\documentclass[aps,prl,reprint, superscriptaddress]{revtex4-2}
\usepackage{blindtext}
\usepackage[colorlinks=true,allcolors=blue]{hyperref}

\usepackage{blindtext}
\usepackage{bm}

\def \beq{\begin{eqnarray}}
\def \eeq{\end{eqnarray}}

\usepackage{amsmath}
\usepackage{amssymb}
\usepackage{graphicx}
\usepackage{graphicx,xcolor}

\graphicspath{{figs}}
\usepackage[capitalize]{cleveref}
\crefname{equation}{Eq.}{}
\crefrangeformat{equation}{Eqs.~(#3#1#4)-(#5#2#6)}

\newcommand{\bqo}{{\hat {\bm q}_0}}

\begin{document}
\title{Non-reciprocal mixtures in suspension: the role of hydrodynamic interactions}

\author{Giulia Pisegna}
\affiliation{Max Planck Institute for Dynamics and Self-Organization (MPI-DS), D-37077 Göttingen, Germany}
\author{Navdeep Rana}
\affiliation{Max Planck Institute for Dynamics and Self-Organization (MPI-DS), D-37077 Göttingen, Germany}
\author{Ramin Golestanian}
\email{ramin.golestanian@ds.mpg.de}
\affiliation{Max Planck Institute for Dynamics and Self-Organization (MPI-DS), D-37077 Göttingen, Germany}
\affiliation{Rudolf Peierls Centre for Theoretical Physics, University of Oxford, Oxford OX1 3PU, United Kingdom}
\author{Suropriya Saha}
\email{suropriya.saha@ds.mpg.de}
\affiliation{Max Planck Institute for Dynamics and Self-Organization (MPI-DS), D-37077 Göttingen, Germany}

\begin{abstract}
The collective chasing dynamics of non-reciprocally coupled densities leads to stable travelling waves which can be mapped to a model for emergent flocking. In this work, we couple the non-reciprocal Cahn-Hilliard model (NRCH) to a fluid to minimally describe scalar active mixtures in a suspension, with the aim to explore the stability of the waves, i.e. the emergent flock in the presence of self-generated fluid flows. We show that the emergent polarity is linearly unstable to perturbations for a specific sign of the active stress recalling instabilities of orientational order in a fluid. Using numerical simulations, we find however that non-reciprocity stabilizes the waves against the linear instability in a large region of the phase space.
\end{abstract}

\maketitle
Life at the micro-scale thrives under conditions of low Reynolds number hydrodynamics~\cite{prandtl:1926,Purcell1977, Happel, Brady_stokes,Najafi2005,lauga2009hydrodynamics,Golestanian2011,Elgeti2015}. Motile and non-motile living agents create persistent long-ranged flows that dominate their interactions, resulting in striking dynamics as exemplified by dancing {\em Volvox} colonies~\cite{DancingVolvox_PhysRevLett.102.168101}, as well as active low Reynolds number turbulence above bacterial carpets \cite{Darnton2004,uchida2010synchronization} and in bacterial suspensions \cite{Wu2000,Wensink2012,Dunkel2013}.
Solvent-mediated interactions have been known to substantially alter the collective behaviour predicted by dry models in a variety of active matter systems~\cite{Gompper2020, marchetti2013hydrodynamics}. For instance, active squirmers generate flows that can suppress motility-induced phase separation \cite{matas2014hydrodynamic, yoshinaga2017hydrodynamic, theers2018clustering,alarcon2013spontaneous}, whereas coupling to active fluids can also stabilize mixtures \cite{Tayar2023}. Moreover, polar and nematic order are known to become unstable due to hydrodynamic interactions \cite{aditi2002hydrodynamic, Shelley_PhysRevLett.100.178103}. While the active stresses generated by polar particles can lead to global spontaneous flows~\cite{voituriez2005spontaneous, tjhung2011nonequilibrium, vogel2013rotation}, in  nematic systems they can lead to turbulence \cite{Giomi_PhysRevX.5.031003,ActiveNematicsDoostmohammadi2018}. The prominent instability can be suppressed via confinement \cite{voituriez2005spontaneous} or background elasticity, and is thought to play a significant role in spindle self-assembly \cite{BruguesNeedlemanSpindle}. Hydrodynamic coupling can also lead to the emergence of order, as seen in the case of coordination between beating cilia and flagella \cite{SynchroReview_Bruot_Cicuta, GITaylor1951, Vilfan_PhysRevLett.96.058102}, which facilitates mass transport \cite{Ling_Kansoo_2024, Wang2022} and leads to the emergence of metachronal waves \cite{uchida2010synchronization,Elgeti2013,Meng2021,Chakrabarti2022}.

\begin{figure}[b]
    \centering
 \includegraphics[width=0.9\linewidth]{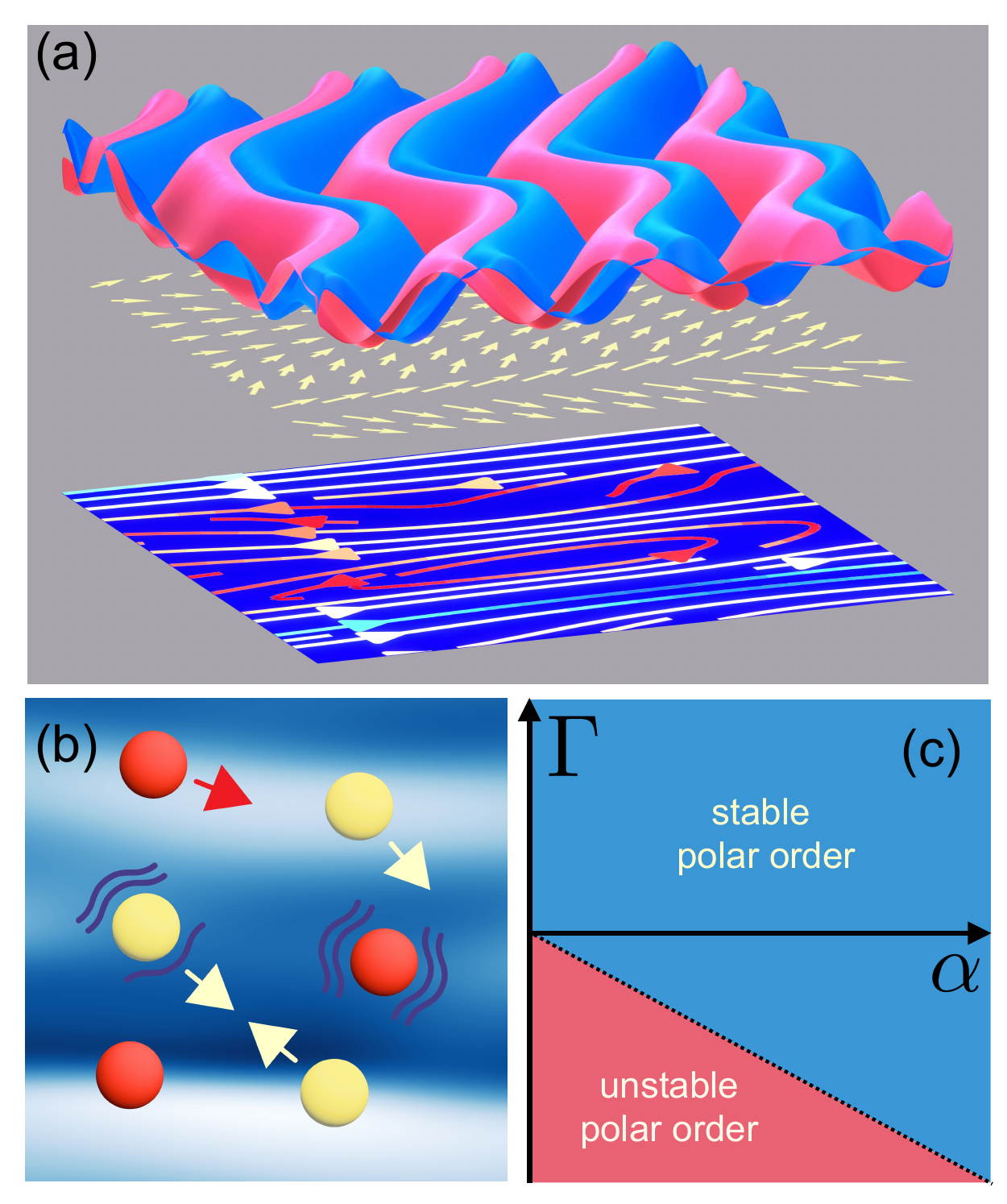}
\caption{Two species of non-reciprocally interacting active particles [depicted in (b)] are represented by the red and blue density profiles in panel (a). Non-reciprocity gives rise to a chasing pattern, whose emergent polar order is quantified by the yellow vector field. Densities are coupled  with a background  fluid [streamlines in panel (a)], which bends the densities layers and the corresponding orientational order. (c) The interplay between the strengths of non-reciprocity $\alpha$ and active stress $\Gamma$ leads to enhanced stability for the polar order (travelling waves).}
\label{fig:drawing}
\end{figure}

Among different manifestations of non-equilibrium activity, a particularly interesting class comprises collective behaviour emerging from action-reaction symmetry breaking due to naturally arising non-reciprocal interactions \cite{soto2014self, Saha2019,agudo2019active}. Non-reciprocity breaks time-reversal symmetry and gives rise to a dynamic chasing behaviour between the species that serves as an effective propulsion mechanism of composite units \cite{soto2014self}. At the collective level, this leads to the emergence of travelling waves, and an accompanying global polar order \cite{saha2020scalar, you2020nonreciprocity, saha2022effervescent}, which are robust to sources of stochasticity in two dimensions and above \cite{pisegna2024emergent}. Many commonly used soft matter systems where the paradigm of non-reciprocity would be applicable, such as mixtures of Janus colloids~\cite{Golestanian2019phoretic} and Quincke rollers~\cite{Bricard2013,Petia_C9SM01163C}, are swimmers in suspensions. A natural question thus arises: how do hydrodynamic interactions impact the emergent travelling waves and the associated polar order that stems from non-reciprocity [see Fig.~\ref{fig:drawing}(a)]?

\begin{figure*}[t]
    \centering
\includegraphics[width=1.00\linewidth]{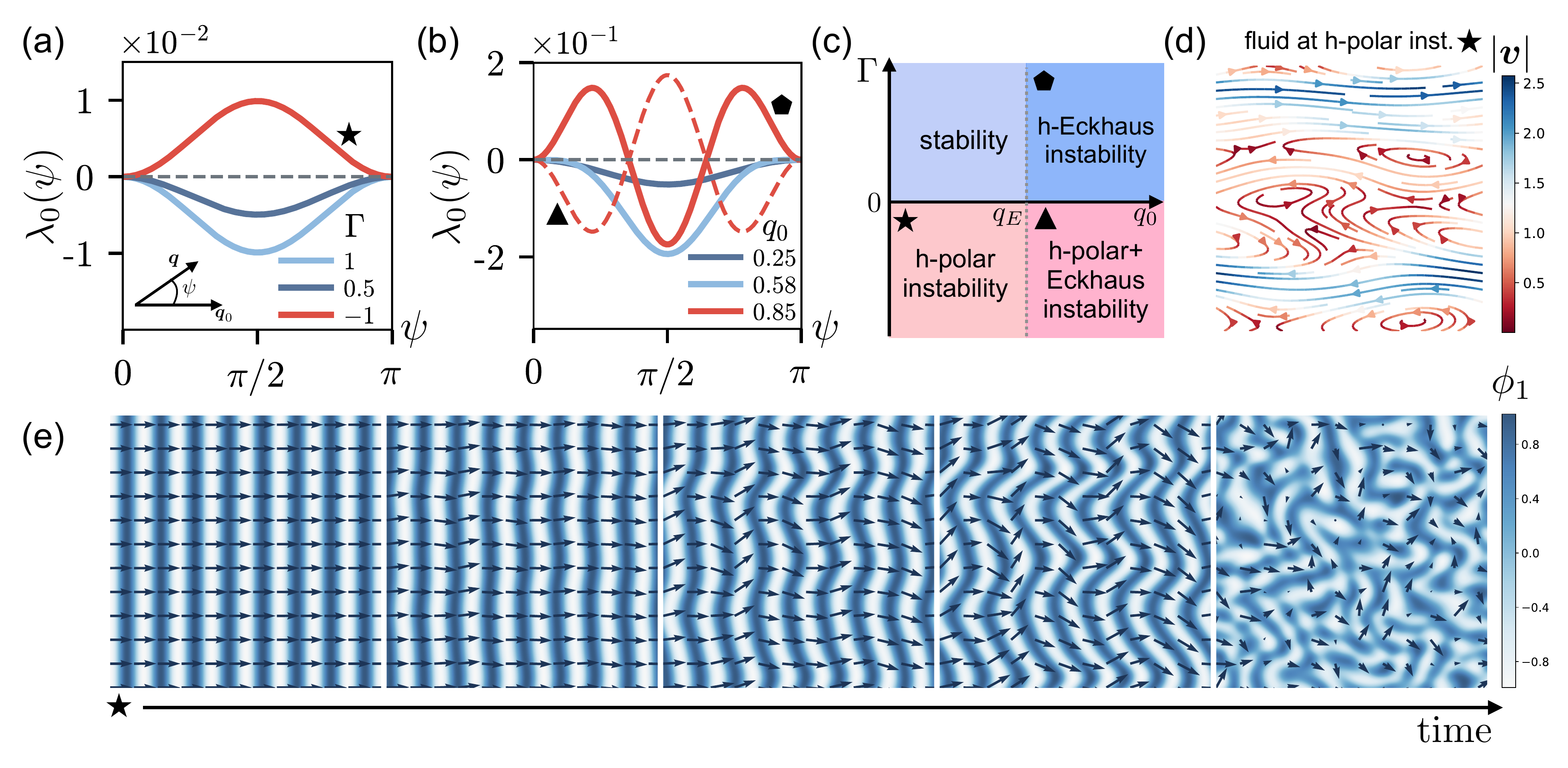}    \caption{{Linear stability of the ordered state.} (a) wave-number independent contribution ($q=0$) of the eigenvalue determining the evolution of the transverse fluctuations. The pattern wave number is fixed at $q_0=0.1$ and we vary the coefficient of the active stress $\Gamma$. (b) Same plot for a fixed $|\Gamma|=1$ (positive for all except the dashed line) and different initial wave-numbers $q_0$. (c) Stability diagram in the $\Gamma-q_0$ plane showing three different routes to linear instability as identified from $\lambda_0(\psi)$. (d) Streamlines of fluid velocity $\bm v$ at the onset of the h-polar instability; the colours indicate the local fluid speed. (e) Time evolution of the density $\phi_1$ undergoing the h-polar instability. The arrows on top represent the corresponding polar order parameter $\bm{J}$. The snapshots are from numerical simulations on a square domain of side $L=32 \pi$ discretized over $512^2$ grid points with $\alpha=1$, $q_0=0.5$, and $\Gamma=-1$. 
}
\label{fig:instability}
\end{figure*}

In this Letter, we address this question by developing a phenomenological model, which we call the hydrodynamic-NRCH (h-NRCH) model for a non-reciprocally interacting binary scalar mixture suspended in a momentum-conserving fluid [see \cref{fig:drawing}(b)]. The h-NRCH model is a modified version of the NRCH model incorporating the effect of solvent-mediated long-range hydrodynamic interactions. 
The central result of our work is that flow structures arising from active stresses lead to a wave-number-independent damping in the dynamics of the (otherwise-) slow modes. The damping is positive for $\Gamma>0$ (a dimensionless scale of the active stress), 
implying that fluctuations decay with a finite timescale in this regime. It changes sign for $\Gamma <0$, thereby triggering an instability reminiscent of the generic instability, which would destroy orientational order in the suspension \cite{aditi2002hydrodynamic}. We uncover another route to instability, a hydrodynamic counterpart of the diffusive Eckhaus instability that arises for any sign of $\Gamma$ for waves with a wavelength below a threshold. These predictions are verified by numerical simulations, where we additionally find that polar order persists in large parts of the parameter space as represented in Fig.~\ref{fig:drawing}(c), in contrast to the results of the linear stability analysis.  

The dynamics of the conserved densities $\phi_{1,2}$ of a non-reciprocally coupled binary mixture of scalar active particles is minimally described by the 
NRCH model \cite{saha2020scalar, you2020nonreciprocity}. The active particles produce active stresses $\mathbb{Q}$ [defined in Eq. \eqref{eq:nrmhsim}] in the fluid in which they are suspended, and they are in turn advected by the resulting hydrodynamic flow fields. The equations of motion for $\phi_a$ ($a \in \{1,2\}$) that satisfy number conservation, and the equations for the Stokesian incompressible velocity field $\bm{v}$  (i.e. $\bm \nabla \cdot \bm v=0$) that enforce momentum conservation in the solvent, define the hydrodynamic NRCH (h-NRCH) model
\begin{eqnarray}§\label{eq:nrmhsim}\begin{aligned}
 & \partial_t \phi_a  + \bm v \cdot \bm  \nabla \phi_a = \nabla^2 \left[  \mu_a - \alpha \varepsilon_{ab} \phi_b \right],\\ 
  &  \eta \nabla^2 \bm{v} = \bm \nabla p + \bm{\nabla} \cdot \mathbb{Q}, \\
  & \mathbb{Q} =  \kappa \left[ \bm{\nabla} \phi_a \bm{\nabla} \phi_a  - \frac{ \mathbb{I}}{d} (\bm{\nabla} \phi_a)^2 \right],
\end{aligned}\end{eqnarray}
where summation over repeated indices is implied. Here, $\varepsilon_{ab}$ is the fully anti-symmetric Levi-Cevita tensor, $\alpha$ is the coefficient of non-reciprocal interaction, and $\eta$ is the viscosity. The pressure $p$ enforces incompressibility, and the chemical potential $\mu_a = \delta F/\delta \phi_a$ is derived from a free-energy $F= \int \mbox{d}^dr \left[-\frac12 (\phi_a \phi_a)+\frac14(\phi_a \phi_a)^2+\frac12(\bm{\nabla} \phi_a)^2\right]$.  Note that we assume the same damping coefficient and stiffness for both species, and absorb them in units of time and space. For $\alpha = 0$ and $\kappa = 1$, Eq.~\eqref{eq:nrmhsim} reduces to the model-H in the Halperin-Hohenberg classification scheme \cite{HH_RevModPhys.49.435}. In contrast, for non-equilibrium conditions, $\kappa$ can take both positive and negative values \cite{tucci2025hydrodynamic}.

For strong non-reciprocity, or large $\alpha$, the system breaks parity and time-reversal symmetry and organizes in stripe-like patterns where one species chases the other. This transition to a novel non-equilibrium phase is easily characterized by the polar order parameter $\bm J\equiv \phi_1 \bm \nabla \phi_2 - \phi_2 \bm \nabla \phi_1$, which measures the local phase difference between the two density fields. The spatial average of $\bm{J}$ vanishes for bulk phase separation \cite{saha2020scalar}, or for defect-riddled states \cite{rana2023defect}, but has a finite value in the moving-phase where $\bm{J}$ is parallel to the wave-vector of the wave \cite{saha2020scalar, rana2023defect, pisegna2024emergent}. Following Ref.~\cite{pisegna2024emergent}, we recast the h-NRCH model \eqref{eq:nrmhsim} in terms of $\bm{J}$ and the amplitude $\rho = \sqrt{\phi_1^2 + \phi_2^2}$. The equation of motion for $\bm J$ is
\begin{align}
\label{eq:wetJ}
  &  \partial_t \bm J +( \bm v + 2 \alpha  \rho^{-2} \bm J) \cdot \bm \nabla \bm J + 2 \alpha \rho^{-2} \bm J (\bm \nabla \cdot \bm J)  \nonumber \\ 
  & = - \mathbb{V} \cdot \bm J  + (r -u |\bm J|^2) |\bm J|^2 \bm J + D_J \bm \nabla (\bm \nabla \cdot \bm J),
\end{align}
where $V_{ij} = \partial_i v_j $ is the velocity gradient tensor. $\bm J$ advects itself with a coefficient proportional to the non-reciprocity $\alpha$, underscoring its similarity to the polar order parameter in flocking dynamics \cite{toner1995long}. The velocity field enters the dynamics in two ways, firstly it advects $\bm J$ [the second term on the L.H.S. of Eq.~\eqref{eq:wetJ}], and secondly through the velocity gradient tensor, which shears (symmetric part) and rotates (anti-symmetric part) $\bm J$ \cite{SM}. The diffusion coefficient $D_J=1- \rho^2- 2 m$ [with $m = \rho^4 (r- u |\bm J|^2)/2$] and the coefficients $r=2(1-\rho^2)/\rho^4$ and $u=2/\rho^8$ of the Mexican-hat potential in the R.H.S. of Eq.~\eqref{eq:wetJ} depend on $\rho$, which evolves as 
\begin{align}
\label{eq:wetrho}
\partial_t \rho + \bm v \cdot \bm \nabla \rho+ \alpha  (\bm \nabla \cdot \bm J)/\rho = m |\bm J|^2/\rho^3 + D_\rho \nabla^2 \rho,
\end{align}
where $D_{\rho} = 5 - 3\rho^2 + 2m$. The active stress in terms of $\bm J$ and $\rho$ is
\beq\label{eq:Q}
 Q_{ij} =  \kappa \left[ \frac{J_i J_j}{\rho^{2}} - \frac{\delta_{ij}}{\rho^{2}d} |\bm J|^2 + \partial_i \rho \partial_j \rho   - \frac{\delta_{ij}}{d}(\bm \nabla \rho)^2 \right].
\eeq
The first two terms in the expression for $\mathbb Q$ are similar in form to the active stress generated in suspensions of active agents irrespective of whether they are polar (and thus self-propelled) or apolar \cite{aditi2002hydrodynamic,saintillan2008instabilities}.

Travelling waves $\phi_{10}= \rho_0 \cos(\bm q_0 \cdot \bm x - \omega_0 t)$ and $\phi_{20}= \rho_0 \sin(\bm q_0 \cdot \bm x - \omega_0 t)$, propagating with angular frequency $\omega_0 = \alpha q_0^2$ and amplitude $\rho_0^2=1- q_0^2$, proposed first for the dry case~\cite{saha2022effervescent,rana2023defect,pisegna2024emergent}, are also exact solutions of the h-NRCH equations in Eq.~\eqref{eq:nrmhsim}. From Eqs.~\eqref{eq:wetJ} and \eqref{eq:wetrho}, we observe that $\bm{v} = 0$, $\bm J = J_0 \hat{\bm q_0}$ and any constant $\rho_0$ are homogeneous solutions of the dynamics. The magnitude $J_0 = \rho_0^2 \sqrt{1-\rho_0^2}$ is determined by the minima of the Mexican-hat potential in Eq.~\eqref{eq:wetJ}, such that $m(J_0,\rho_0) = 0$. 

To test the stability of the ordered state, we perturb the fields around their stationary values, substituting $\rho(\bm x,t) = \rho_0 + \delta \rho (\bm x,t)$ and $\bm J (\bm x,t) =  J_0 \hat{\bm q_0} + \delta J_{\parallel}(\bm x,t) \hat{\bm q_0} + \delta \bm J_{\perp}$ in Eqs.~\eqref{eq:wetJ} and \eqref{eq:wetrho}, and probe the evolution of the fluctuations in space and time. For the vector order parameter $\bm J$, we have decomposed the fluctuations 
as parallel and perpendicular with respect to the direction of propagation, i.e. $\delta \bm J_{\perp} \cdot  \hat{\bm q}_0 = 0$. We retain terms until linear order in the perturbations to obtain the dynamical matrix for $\delta \bm{J}_{\perp}$ and $\delta \rho$ whose eigenvalues determine the stability of the polar order. To reach this point we eliminate the pressure field using the condition of incompressibility and express $\bm v$, which we likewise decompose into parallel and perpendicular components, in Fourier space in terms of $\delta \bm J_\perp$ and $\delta \rho$ \cite{SM}. To the lowest order, we find 
\beq\label{eq:devsStress}
\delta v_\parallel = - \frac{i\Gamma q_0}{q} \left[ \frac{ (\bm q \cdot \delta \bm J_\perp)}{q} + 2 q_0 \rho_0 \cos \psi  \sin^2 \psi \;\delta \rho  \right],
\eeq 
where $\psi$ is the angle between $\bm q$ and $\bm q_0$, namely, $\psi=\cos^{-1}\left({\hat {\bm q}} \cdot \bqo\right)$, 
and $\Gamma = \kappa/\eta$ is the only relevant dimensionless parameter. 
In $d=2$, we can readily substitute this expression in the  dynamics of \eqref{eq:wetJ} and \eqref{eq:wetrho} and focus on the modes $\delta J_\perp (\bm q,t)$ whose linearized dynamics reads 
\begin{equation}\label{eq:J}
\partial_t \delta J_\perp = \left[  \lambda_0(\psi) - i q v_g(\psi) \cos \psi - q^2 D(\psi)  \right] \delta J_{\perp}.
\end{equation}
Here, $\lambda_0$ is a wave-number-independent growth rate given as 
\begin{equation}
\lambda_{0}(\psi) = - \Gamma q_0^2 \rho_0^2 \sin^2 \psi \left( \gamma  + \frac{2 q_0^2}{\rho_0^2} \sin^2 \psi \right), 
\label{eq:eigenunstable}
\end{equation}
where
\begin{equation}
    \gamma = \frac{1-3 q_0^2}{1-q_0^2}.
    \label{eq:gamma-def}
\end{equation}
The generation of a wave-number-independent and anisotropic growth rate $\lambda_0$ is a key result of this work. $\lambda_0$ arises either from the advection of scalar densities by self-generated long-range hydrodynamic flows or, analogously, from shear stresses deforming the polarity. The perturbations propagate with the velocity $v_g$ and diffuse with the coefficient $D$, both of which are anisotropic for $\Gamma \neq 0$. $D$ stabilizes the instability that arises at zero wave-number due to $\lambda_0$ at small enough system sizes; see the Appendix for the expressions for $v_g$ and $D$, and \cite{SM} for further discussion.
If the sign of $\lambda_0$ is positive for any $\psi$, fluctuations in that direction will grow exponentially and eventually destroy the ordered state. From Eq.~\eqref{eq:eigenunstable}, we note that the sign of $\lambda_0$ depends on $\psi$, and on the wave-number of the reference state $q_0$, especially through $\gamma$. This quantity changes sign from positive to negative at $q_0 = q_{\rm E} = 1/\sqrt{3}$, a threshold that matches the Eckhaus instability criterion for layered densities \cite{cross2009pattern}.  

We will now discuss the impact of all four possible combinations of the sign of $\Gamma$ and $\gamma$ on the stability of the emergent order. For $q_0<q_{\rm E}$, $\gamma$ is positive and the sign of $\lambda_0$ is simply opposite to that of $\Gamma$. For $\Gamma>0$, $\lambda_0$ is negative ensuring the existence of a region of parameter space where travelling waves persist even when coupled to a Stokesian fluid. Conversely, for $\Gamma <0$ orientational order is linearly unstable and the system transits to a new non-equilibrium steady state. As the instability destroys the emergent flock, we call the instability in this sector, the h-polar instability; see Fig.~\ref{fig:instability}(c). Yet another route to an instability appears when $\gamma$ reverses sign signaling an instability for $\psi$ around zero for $\Gamma>0$ and around $\psi = \pi/2$ for $\Gamma<0$; see Fig.~\ref{fig:instability}(b). We call these the h-Eckhaus and the h-polar+Eckhaus instabilities. The standard diffusive Eckhaus instability of a patterned density happens due to the advection of amplitude fluctuations in the direction of propagation. The hydrodynamic Eckhaus instability occurs due to the same mechanism entering via the second term in the $v_{\parallel}$ in Eq.~\eqref{eq:devsStress}.

To confirm our analytical predictions we run numerical simulations of the h-NRCH Eq.~\eqref{eq:nrmhsim} using parameters where we expect travelling waves for $\Gamma = 0$ \cite{rana2023defect}, and explore all four regimes of Fig.~\ref{fig:instability}(c). All simulations are initialized with $\bm{v}=0$ and a travelling wave, i.e. $\phi_{a} = \phi_{a0}$, with random uniform perturbations \cite{SM}. 
In the stable regime, the travelling waves are stable and perturbations decay away in finite time. In the region where the h-polar instability is expected, for large enough $|\Gamma|$, the waves are destroyed and the steady state is one of structure-less spatiotemporal chaos. The snapshots of Fig.~\ref{fig:instability}(e) show the evolution of the layered structure at the onset of the instability, where it is apparent that the hydrodynamic instability happens through deformation of the densities layers, see SM~\cite{SM} for a quantification of the large deformations. These pictures are reminiscent of the buckling instability of smectic A under dilative stress \cite{ribotta1977mechanical, oswald1982undulation} with the important difference that the mechanical stresses are generated by the fluid as shear flows along the ordering direction. This pulls the layers apart finally disrupting them; see Fig.~\ref{fig:instability}(d,e). For $q_0 > q_{\rm E}$ the wave is destroyed due to both the hydrodynamic [see Eq.~\eqref{eq:eigenunstable}] and the conserved Eckhaus instabilities that appears in the diffusion coefficient $D$ [see Eq.~\eqref{eq:J}]. The final state of the system is determined by the sign of $\Gamma$. For $\Gamma>0$, the upper left quadrant in Fig.~\ref{fig:instability}(c), the travelling waves are reformed with a wavelength smaller than $q_{\rm E}$. For $\Gamma<0$, we observe spatiotemporal chaos at long times, without prominent length or time scales \cite{SM}.

\begin{figure}
    \centering
\includegraphics[width=\linewidth]{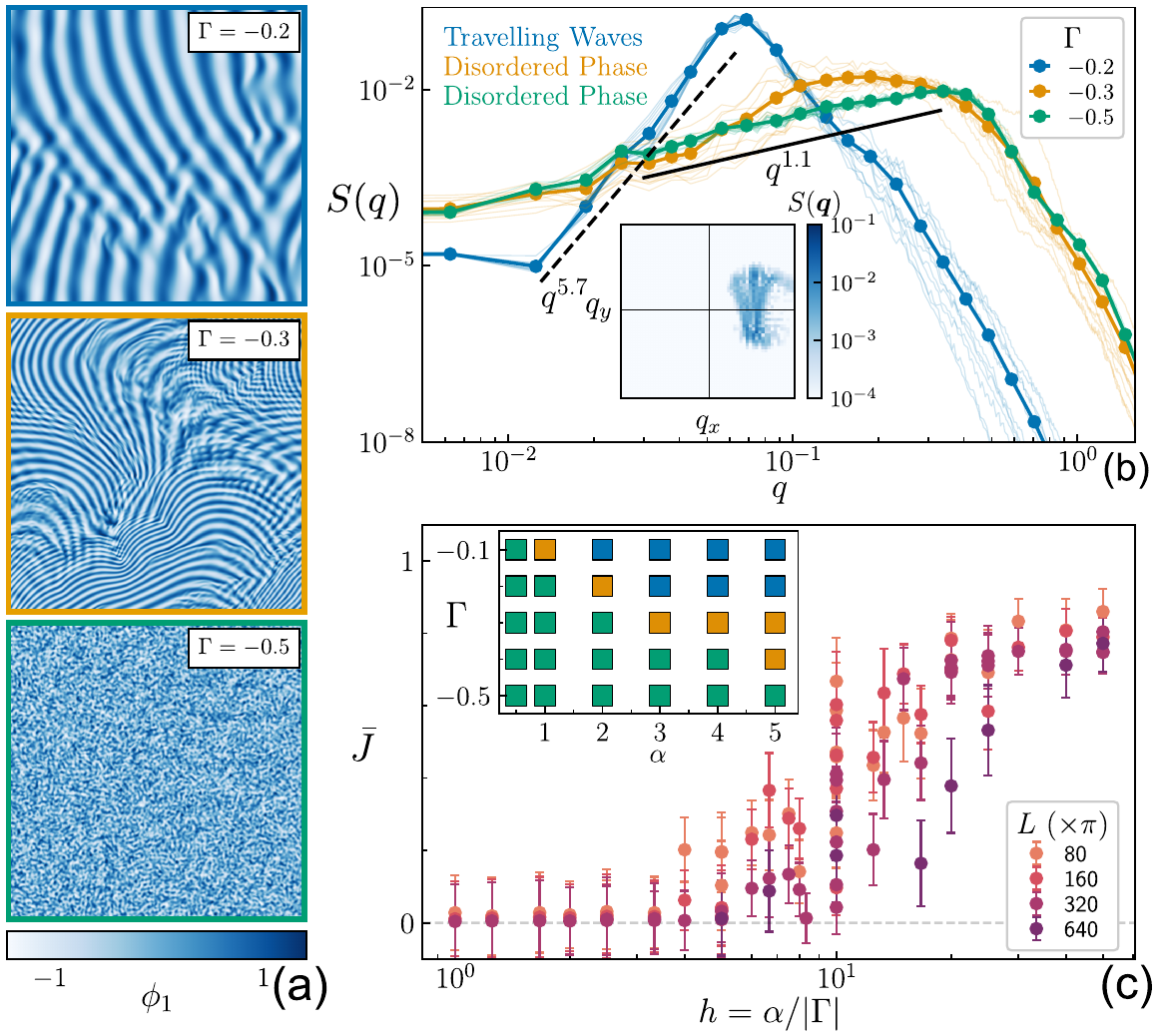}
    \caption{Order and disorder in the full nonlinear theory.  (a) The snapshots show the two distinct phases we observe in the $\alpha-\Gamma$ parameter space: Travelling Waves (blue), and Disordered Phase, as shown by snapshots in the transitional state (orange) and deep in the disordered phase (green).
   (b) Shell-averaged structure factor $S(q)$ in different states for a fixed $\alpha$ and different $\Gamma$ marked by the same colours as the borders. For travelling bands, $S(q)$ shows a peak at the dominant wave-number. For the crossover and disordered states, we observe a large $\sim q^1$ power law, with an additional contribution at larger wave-numbers for the transitional state. In both (a) and (b), we show the data for $\alpha=4$. Inset: In the polar phase, structure factor $S(\bm q)$ is anisotropic, consistent with the travelling bands. (c) Average polar order $\bm J$ versus $h = \alpha/|\Gamma|$ shows a disorder-order transition with increasing $h$ for all system sizes. Inset : Phase diagram in the $\alpha-\Gamma$ space for $L=320\pi$. }
    \label{fig:phase-diagram}
\end{figure}

So far in the paper, the role played by $\alpha$ has been to determine the reference state for the stability analysis. Focusing on $\Gamma < 0$, we now show that $\alpha$ plays a pivotal role in stabilising the polar order that it creates. We perform simulations for varying $\alpha$, $\Gamma$, and the box size $L$, to obtain the steady state phase diagram summarized in \cref{fig:phase-diagram}(c). We find two distinct categories of dynamics: a travelling band phase and a disordered phase, which contains a transitional region consisting of well defined density layers with many dislocations, shown in representative snapshots of field $\phi_1$ in Fig.~\ref{fig:phase-diagram}(a). The phases are characterized by analysing the circularly averaged structure factor $S(q)$ \cite{SM}; see Fig.~\ref{fig:phase-diagram}(b) and the average polar order $\bm J$ in the steady state in Fig.~\ref{fig:phase-diagram}(c). For the disordered phase, $S(q,t) \sim q^{1+\delta}$ for $q < q_m$ with $\delta \ll 1$, which implies that the order parameter is uncorrelated at large length scales. Here $q_m \lesssim 1$ is the wave-number at which $S(q)$ peaks. Consequently, the dissipation spectrum ${\cal E}(q,t)$ behaves similarly to that of a randomly-stirred Stokes fluid and we obtain ${\cal E}(q,t) \sim q^{-1}$ for $q < q_m$ \cite{SM}. $S(q)$ for the different cases are shown in Fig.~\ref{fig:phase-diagram} (b), from which we confirm that the spectral properties of the transitional and strongly-disordered cases are similar. Therefore, we observe that within the linearly unstable region, global polar order persists for $|\alpha|\gg|\Gamma|$, as we will discuss below. 


For a given $\Gamma$, there exists a critical $\alpha_c(L)$ where we observe the transition from the disordered phase to the travelling bands. From our simulations, we observe that $\alpha_c$ increases with $\Gamma$, thus, we assume the simple scaling relation $\alpha_c \sim |\Gamma|$ to quantify the transition in terms of rescaled parameter $h=\alpha/\alpha_c \sim \alpha/|\Gamma|$ \footnote{Note that this might not be the correct scaling relation, in general we expect $\alpha_c \sim |\Gamma|^a$, where $a$ is some positive number.}. In \cref{fig:phase-diagram}(c), we plot the average polar order $\bar{J} \equiv |\langle \bm J  \rangle| /J_0 $ versus $h$ for different system sizes. The magnitude of $\bm{J}$ depends on the dominant length scales of the polar patterns and we normalize it by $J_0$ which is estimated from $S(q)$ \cite{SM}. This ensures that for a perfect travelling wave, $ \bar{J} = 1$ irrespective of the wave-number, while $\bar{J} = 0$ for the disordered case. Consistent with the snapshots, we find a disorder to order transition with increasing $h$ for all values of $L$. $\bar{J}$ vanishes for disordered states ($h \sim 1$), and in the transitional state ($h \lesssim h_c$), we find noisy partial order. For $h > h_c$, global polar order sets in and $\bar{J} \sim 1$ is close to it's expected value for perfect polar order. Thus, even in the presence of a small $q$ linear instability, the system shows global polar order for $h>h_c$. From our simulations, we infer $h_c \sim 10$.

For a binary non-reciprocal mixture it is useful to rewrite the dynamics of the fields as the coupled dynamics for a vector order parameter and an amplitude~\cite{pisegna2024emergent}. With this transformation, the hydrodynamic stress resembles the active stress that appears in theories for flocking in a fluid~\cite{aditi2002hydrodynamic, saintillan2008instabilities}. Fluctuations around the ordered state drive velocity fluctuations that shear the polarity to either stabilize it, or destabilize it to produce feature-less spatiotemporal chaos. The order-disorder transition in a suspension is \emph{stabilized} by non-reciprocity and a new state with multiple dislocations appears in the regime where the transition happens. The instabilities of the h-NRCH model bear similarities to the generic instability in active matter as regards the form of the active stress term, and the mechanism of the instability leading to the generation of a wave-number independent damping. The important difference is that h-polar instability is not generic as it does not destabilize the polar order for both signs of $\Gamma$ in the Eckhaus stable regime $q<q_{\rm E}$. Our findings highlight the complex interplay between non-reciprocal interactions and hydrodynamics in determining collective behaviour in active systems and offers new insights into pattern formation in a suspension. In future work, we plan to explore the role of geometric confinement \cite{maitra2018nonequilibrium}, or inertia~\cite{Inertia_PhysRevX.11.031063, Rana2024Feb} on the emergent polar order. Our study highlights the stabilizing effect of non-reciprocity, in line with recent reports of similar stabilizing effects in other contexts \cite{rana2023defect,parkavousi2024enhancedstabilitychaoticcondensates}. 

\acknowledgements
We acknowledge discussions with Sriram Ramaswamy, Ananyo Maitra, and Gennaro Tucci. This work has received support from the Max Planck School Matter to Life and the MaxSyn-Bio Consortium, which are jointly funded by the Federal Ministry of Education and Research (BMBF) of Germany, and the Max Planck Society.

 
\textit{Appendix: group velocity and effective diffusion coefficient.} Here we report the explicit expressions for the group velocity, and the diffusion coefficient for $\delta \bm J_{\perp}$, see Eq.~\eqref{eq:J}. These expression are derived assuming small $\Gamma$, for details see the SM~\cite{SM}. For $\Gamma<2$ we find that the phase propagates anisotropically with the modified speed $v_g(\psi)$ given by
\beq 
v_g(\psi) = q_0 \alpha \left[  2 - \Gamma \sin^2 \psi \left(\gamma+ \frac{2 q_0^2}{\rho_0^2} \sin^2 \psi \right)\right].
\eeq 
The anisotropic diffusion coefficient is,
\begin{align}\label{eq:fullD} 
    D(\psi) &= \gamma q_0^2 \cos^2 \psi + q_0^2 \sin^2 \psi + D_m(\psi),
\end{align}
where $D_m$ is a cross-diffusion term (i.e. it vanishes for $\psi  = 0, \pi/2$) proportional to $\Gamma$ generated by the coupling to the fluid. The Eckhaus factor $\gamma$ defined in Eq.~\eqref{eq:gamma-def} determines the sign of the longitudinal diffusivity $D(0)$ in Eq.~\eqref{eq:fullD}. This triggers a conserved version of the instability that occurs in the complex Landau-Ginzburg dynamics~\cite{aranson2002world} and appears in NRCH without the fluid ~\cite{saha2022effervescent, pisegna2024emergent}. The full expression for $D_m$ is
\begin{align}\label{eq:Dm}    
    D_m(\psi) &= -\frac{\Gamma}{4 \rho_0^2} \sin^2(2 \psi) \bigg[ \alpha^2 \left(\gamma + 2 \frac{q_0^2}{\rho_0^2} \sin^2 \psi \right) \nonumber \\
    &\quad + q_0^2 \left( 2 + 4 q_0^2 \sin^2 \psi - 7 q_0^2 \right) \bigg].
\end{align}
For $\Gamma<0$, $D_m$ is positive in large parts of the parameter space. The diffusive instability appearing in this regime due to a change of sign of $D_m$ is also controlled by $\gamma$. Notice that the term proportional to $\alpha^2 \Gamma$ in Eq.~\eqref{eq:Dm} is the same as the one that appears in $\lambda_0$ in Eq.~\eqref{eq:eigenunstable}. 

Where the h-polar instability occurs, $D$ determines the threshold system size above which the h-polar instability sets in, see SM~\cite{SM} for details. It competes with $\lambda_0$ in determining the system size $L_I = (\sqrt{D /\lambda_0})$ below which the h-polar instability is suppressed. From Fig.~\ref{fig:instability} (a), $\lambda_0$ is maximum at $\psi = \pi/2$, an estimate for $L_I \approx {2 \pi}/|\Gamma| \rho_0^2$, i.e. larger the coupling $|\Gamma|$, the smaller is the length-scale for which the instability becomes manifest. For perturbations along $\psi=\pi/2+\epsilon$, $\alpha$ dependent $D_m$ determines $L_I$ where we find that for a given $\Gamma$, $L_I$ increases rapidly with $\alpha$ implying that it contributes to stabilize the flock; see SM~\cite{SM} for a detailed report.

\bibliographystyle{apsrev4-2}  
\bibliography{biblio_NRCH} 
\end{document}